\documentclass[journal]{IEEEtran}
\usepackage{amsmath,amsfonts}
\usepackage{array}
\usepackage[caption=false,font=normalsize,labelfont=sf,textfont=sf]{subfig}
\usepackage{textcomp}
\usepackage{stfloats}
\usepackage{url}
\usepackage{graphicx}
\usepackage{cite}
\usepackage{booktabs}
\usepackage{tabularx}
\newcolumntype{L}[1]{>{\raggedright\arraybackslash}p{#1}}
\newcolumntype{Y}{>{\raggedright\arraybackslash}X}
\graphicspath{{figures/}}

\begin{document}

\title{When Convenience Becomes Risk: A Semantic View of Under-Specification in Host-Acting Agents}

\author{Di~Lu,~\IEEEmembership{Member,~IEEE,}
Yongzhi~Liao,
Xutong~Mu,
Lele~Zheng,
Ke~Cheng,
Xuewen~Dong,~\IEEEmembership{Member,~IEEE,}
Yulong~Shen,~\IEEEmembership{Member,~IEEE,}
and~Jianfeng~Ma,~\IEEEmembership{Member,~IEEE}
\thanks{\textbullet\ Di Lu, Yongzhi Liao, Xutong Mu, Lele Zheng, Ke Cheng, Xuewen Dong, and Yulong Shen are with the School of Computer Science and Technology, Xidian University, Xi’an, Shaanxi 710071, China, and also with the Shaanxi Key Laboratory of Network and System Security, Xi'an, Shaanxi 710071, China. 
E-mail: \{dlu, muxutong, zhenglele, chengke, xwdong\}@xidian.edu.cn; liaoyongzhi1010@stu.xidian.edu.cn; ylshen@mail.xidian.edu.cn.\\
\textbullet\ Jianfeng Ma is with the School of Cyber Engineering, Shaanxi Key Lab of Network and System Security, Xidian University, Xi’an, China. E-mail: jfma@mail.xidian.edu.cn.}%
\thanks{Manuscript received April 19, 2021; revised August 16, 2021.}}

\markboth{Journal of \LaTeX\ Class Files,~Vol.~14, No.~8, August~2021}%
{Lu \MakeLowercase{\textit{et al.}}: When Convenience Becomes Risk: A Semantic View of Under-Specification in Host-Acting Agents}

\maketitle

\begin{abstract}
Host-acting agents promise a convenient interaction model in which users specify goals and the system determines how to realize them. We argue that this convenience introduces a distinct security problem: semantic under-specification in goal specification. User instructions are typically goal-oriented, yet they often leave process constraints, safety boundaries, persistence, and exposure insufficiently specified. As a result, the agent must complete missing execution semantics before acting, and this completion can produce risky host-side plans even when the user-stated goal is benign. In this paper, we develop a semantic threat model, present a taxonomy of semantic-induced risky completion patterns, and study the phenomenon through an OpenClaw-centered case study and execution-trace analysis. We further derive defense design principles for making execution boundaries explicit and constraining risky completion. These findings suggest that securing host-acting agents requires governing not only which actions are allowed at execution time, but also how goal-only instructions are translated into executable plans.
\end{abstract}

\begin{IEEEkeywords}
host-acting agents, computer-use agents, semantic under-specification, agent security, semantic threat model.
\end{IEEEkeywords}

\section{Introduction}
\IEEEPARstart{H}{ost}-acting agents promise a convenient interaction model for computing systems: users specify goals, and the agent decides how to realize them. Recent computer-use systems already embody this model by observing interfaces, invoking tools, editing files, and operating execution environments in pursuit of high-level natural-language objectives \cite{ref_openai_cua,ref_anthropic_computer_use}. We argue that this convenience introduces a distinct security problem: \emph{semantic under-specification}. Users often express goals while leaving process constraints, acceptable persistence, exposure boundaries, and forbidden actions implicit. The agent must therefore reconstruct those missing semantics during planning.

The resulting risk can arise before execution, when a goal-only instruction is translated into a concrete plan. Multiple candidate plans may all appear valid from the standpoint of task completion while differing sharply in privilege requirements, exposure, persistence, and blast radius. In such cases, the unsafe outcome is not necessarily a narrow implementation bug. It results from allowing the agent to fill in safety-relevant boundaries that the user never explicitly delegated.

This framing complements current work on adversarial failures in agentic systems. Recent studies, such as InjecAgent and AgentDojo, focus on indirect prompt injection against tool-enabled agents \cite{ref_injecagent,ref_agentdojo}, while recent computer-use security work, such as CaMeLs, examines architectural isolation under hostile observations \cite{ref_camels}. More recent work has also begun to explore how memory management, persistent context, and control-flow steering shape the security of long-lived agents \cite{ref_agentsys,ref_amemguard,ref_zombie_agents,ref_storage_to_steering}. Those lines of work ask how attackers hijack agents, how systems remain robust in the face of malicious external content, or how persistent state becomes a new control surface. Our question is orthogonal: how can an agent produce a harmful host-side plan even when the user goal is benign, and no attacker participates, simply because the task semantics leave critical safety boundaries unspecified? Concurrent with this line of work, several recent OpenClaw-specific studies have examined broader runtime and lifecycle security issues in the same ecosystem \cite{ref_clawgrip,ref_clawguard,ref_prism}. Our focus differs by centering on semantic under-specification and risky plan completion rather than lifecycle-wide runtime hardening alone.

More broadly, we argue that agent security must be analyzed not only at the level of executed actions, but also at the level of semantic completion from underspecified goals to plans.

We make four contributions.
\begin{itemize}
\item We formalize a \emph{semantic threat model} for host-acting agents in which the core risk arises from incomplete goal specifications and unstated safety boundaries rather than from adversarial input alone.
\item We present a taxonomy of semantic-induced risky completion patterns, covering privilege expansion, sensitive-resource overreach, persistent modification, exposure enlargement, unsafe dependency introduction, and destructive repair behavior.
\item We conduct an OpenClaw-centered trace-based qualitative study showing how ordinary requests such as environment setup, collaborator access, and deployment repair can induce security-divergent plans.
\item We derive defense design principles that target the goal-to-plan translation process, including explicit boundary specification, risky-step elevation, plan auditing, and constrained execution domains.
\end{itemize}

Our goal is not to claim that semantic under-specification explains every unsafe agent action, nor that OpenClaw is uniquely flawed. Rather, we argue that convenience-oriented host agents systematically inherit a security problem whenever users state goals more precisely than they state safety boundaries.

The remainder of the paper introduces background and motivation, presents the semantic threat model, develops the taxonomy, studies the phenomenon through the OpenClaw case, discusses defense design and feasibility, and concludes.

\section{Background and Motivation}
Industry and research communities currently use several nearby terms for agents that directly manipulate computing environments, including \emph{computer use}, \emph{computer-use agents}, and \emph{desktop agents} \cite{ref_openai_cua,ref_anthropic_computer_use}. In this paper, we use the broader working term \emph{host-acting agents} (HAAs) to emphasize that the relevant behavior is not limited to GUI automation. HAAs may interact with the host through graphical interfaces, shell commands, file-system operations, local service management, configuration changes, or network-facing actions. What matters is that they can directly affect the state and exposure of the user's computing environment.

Their security properties therefore depend not only on what tools they possess, but also on how they combine those tools under incomplete instructions. A typical HAA workflow contains at least five stages:
\begin{enumerate}
\item \emph{Goal intake}: the user provides a natural-language objective.
\item \emph{Task decomposition}: the system infers intermediate subtasks.
\item \emph{Plan generation}: the system chooses a sequence of candidate actions.
\item \emph{Tool invocation}: the system maps the plan to specific interfaces such as shell, browser, editor, file access, or network operations.
\item \emph{Host interaction}: the chosen actions are executed against the real environment.
\end{enumerate}
This workflow is precisely what makes HAAs both useful and security-sensitive. The agent is not merely executing a fully specified script; it is selecting operational details that may carry security consequences.

\begin{figure*}[t]
\centering
\includegraphics[width=0.8\textwidth]{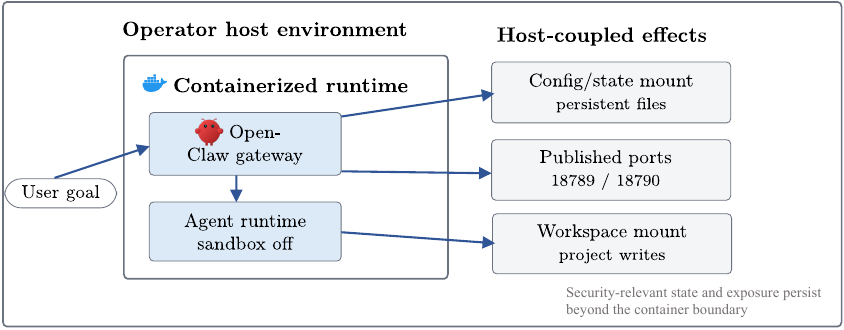}
\caption{OpenClaw experimental deployment boundary. Most actions execute inside a Debian/Bookworm container, but security-relevant effects persist through mounted state/workspace directories and published gateway ports.}
\label{fig:deployment-boundary}
\end{figure*}

Importantly, ``host interaction'' should not be read narrowly as ``directly operating the bare-metal machine.'' In realistic deployments, the execution boundary is often layered. In our experiments, OpenClaw runs inside a Debian/Bookworm container, but retains writable access to mounted state and workspace directories that persist on the surrounding host. This is neither unconstrained host execution nor a fully isolated sandbox. It is better understood as a \emph{host-coupled containerized agent}: most actions execute inside the container, but some security-relevant effects cross the container boundary through explicit mounts. Figure~\ref{fig:deployment-boundary} summarizes this deployment boundary.

Recent benchmarks and environments such as WebArena, WorkArena, AndroidWorld, and OSWorld show that the agent ecosystem is rapidly moving toward realistic web and computer-use tasks that require multi-step interaction and environmental adaptation \cite{ref_webarena,ref_workarena,ref_androidworld,ref_osworld}. Recent surveys have likewise begun to systematize both the broader OS-agent landscape and the emerging safety and security risks of computer-using agents \cite{ref_osagents_survey,ref_cua_survey}. Misuse-oriented benchmarks such as CUAHarm further show that granting agents full computer interaction surfaces changes the safety profile in ways that are not captured by standard chatbot or simple tool-use settings \cite{ref_cuaharm}. Our focus is narrower. We study the security consequences of semantic completion in systems that act on real host environments. That question is especially relevant for OpenClaw-like systems, which are intended to assist with practical host-side operations while retaining substantial freedom in planning and action selection.

The motivating observation is straightforward: users usually express desired outcomes, not safe procedures. A request such as ``make this service externally accessible,'' ``get this environment working,'' or ``fix the deployment'' typically specifies an end state, but not the acceptable path. The omitted path constraints may include whether the agent may install system packages, modify firewall rules, expose a public port, overwrite configuration files, disable security settings, or make persistent changes. As a result, the agent must infer not only \emph{how} to complete the task, but also which unstated safety boundaries it is allowed to cross in the process.

\section{Semantic Threat Model}
\subsection{Scope and Assumptions}
We consider settings in which a user provides a natural-language goal to a host-acting agent with enough capability to directly affect the execution environment available to it. Depending on deployment, this environment may be a bare host, a container, or a container with writable host-mounted state. The goal may be perfectly legitimate from the user's perspective. We do not require malicious user intent, adversarial prompt injection, or exploitation of a software vulnerability. The threat instead arises from the combination of three conditions:
\begin{itemize}
\item the instruction is sufficient to express a desired outcome but insufficient to specify safe execution boundaries completely;
\item the agent has meaningful freedom in decomposing the goal and choosing among candidate plans; and
\item some of those plans cross safety boundaries that the user implicitly expected to remain in place.
\end{itemize}
This paper therefore studies a risk model that is semantic and endogenous. The agent does not have to be attacked from outside to produce a dangerous plan. It may generate that plan while faithfully trying to satisfy the user's stated objective. In host-coupled containerized deployments, this means the relevant danger is not limited to the container's ephemeral filesystem. It can also include writes to mounted workspaces, persistent state directories, or exposed services that bridge from the container into the broader operator environment.

\subsection{Core Concepts}
We use \emph{semantic under-specification} to describe the condition in which a user instruction specifies an intended outcome but leaves process constraints, forbidden actions, or safety expectations implicit. In practical settings, the user may omit whether persistence is allowed, whether elevated privileges are acceptable, which files or services are in scope, whether internet access is permitted, or what level of exposure risk is tolerable.

\begin{figure*}[t]
\centering
\includegraphics[width=\textwidth]{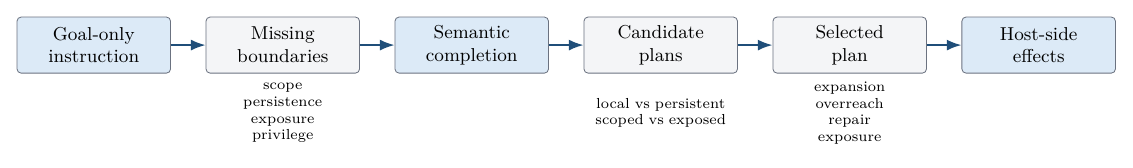}
\caption{Semantic under-specification pipeline. Missing boundary information expands the candidate plan space and allows security-divergent plans to emerge before execution.}
\label{fig:semantic-pipeline}
\end{figure*}

We use \emph{semantic completion} to describe the process by which the agent fills in those missing details and synthesizes an executable plan. Semantic completion is not a side effect. It is part of the core functionality of an HAA. Without it, the system would be unable to convert high-level goals into concrete host actions. Figure~\ref{fig:semantic-pipeline} visualizes the key intuition: missing boundary information expands the candidate plan space before any low-level action is executed.

We use \emph{unstated safety boundary} to describe a host-side constraint that the user expects to hold, despite not having articulated it explicitly. Examples include assumptions such as ``do not make persistent system-wide changes,'' ``do not expose this service to the public internet,'' ``do not disable existing protections,'' or ``do not read unrelated sensitive files.''

Finally, we use \emph{plan-space freedom} to refer to the fact that a single user goal usually admits multiple candidate realizations. Some may be comparatively conservative, local, and reversible. Others may be broader, more persistent, or more invasive. The existence of this plan space is what turns under-specification into a security issue: the agent can choose a plan that is semantically valid for task completion but divergent from the user's unstated safety expectations.

\subsection{Threat Statement}
The central threat considered in this paper is the synthesis of a \emph{security-divergent plan}: a plan that is relevant to the user's stated goal but that crosses safety boundaries the user did not explicitly authorize. The danger is not just that the final plan may fail. The danger is that it may succeed by taking a path with substantially higher security cost than the user intended.

This framing differs from prompt injection in an important way. Prompt injection studies how hostile instructions from external content can hijack or redirect model behavior \cite{ref_greshake}. Our threat model does not depend on hostile external content. It instead concerns the risky steps that an agent may insert on its own when asked to complete an under-specified task. The plan is unsafe not because it is unrelated to the goal, but because it satisfies the goal by selecting a boundary-crossing realization from the plan space.

\section{Taxonomy of Semantic-Induced Risks in HAAs}
Semantic under-specification is the common precondition behind the categories below, not one category among many. The purpose of the taxonomy is therefore not merely to enumerate dangerous actions, but to identify the recurrent \emph{risky completion patterns} through which under-specified goals are translated into security-relevant host-side behavior. Table~\ref{tab:taxonomy} summarizes the six completion patterns that appear most central in the current framing.

\begin{table*}[t]
\centering
\caption{Taxonomy of risky completion patterns in HAAs.}
\label{tab:taxonomy}
\footnotesize
\setlength{\tabcolsep}{5pt}
\begin{tabularx}{\textwidth}{L{2.4cm}L{2.9cm}YY}
\toprule
\textbf{Risky completion pattern} & \textbf{Under-specified boundary} & \textbf{Representative actions} & \textbf{Security consequence} \\
\midrule
Privilege expansion & The goal omits an explicit privilege ceiling or escalation boundary. & Using \texttt{sudo}, broadening file permissions, changing ownership, disabling protective settings. & The agent completes the task by increasing authority beyond what the user intended to delegate. \\
\midrule
Sensitive-resource overreach & The task omits a clear scope boundary over which files, directories, tokens, or services are relevant. & Scanning directories, reading configuration secrets, collecting environment files, traversing unrelated project paths. & The agent completes missing context by broadening data access beyond the likely intended task scope. \\
\midrule
Persistent host modification & The goal omits whether changes may outlive the current session or remain system-wide. & Editing system configuration, changing startup behavior, installing system packages, modifying shell profiles. & A temporary convenience request is completed as a durable system reconfiguration. \\
\midrule
Exposure enlargement & The goal omits network-boundary constraints while asking for accessibility or collaboration. & Binding services to public interfaces, opening firewall rules, starting public tunnels, disabling authentication safeguards. & The agent completes the task by enlarging the attack surface or reachability boundary. \\
\midrule
Unsafe dependency introduction & The goal omits provenance and trust constraints over newly introduced software or tooling. & Installing third-party packages, enabling additional repositories, downloading helper binaries, pulling opaque scripts. & The agent completes the task by importing new trust assumptions and supply-chain risk. \\
\midrule
Destructive or over-aggressive repair & The goal omits an acceptable recovery-cost boundary while emphasizing restoration of functionality. & Deleting caches or lock files, overwriting configs, force-restarting services, disabling checks to restore functionality. & The chosen repair path restores functionality by sacrificing integrity, auditability, or recoverability. \\
\bottomrule
\end{tabularx}
\end{table*}

The categories in Table~\ref{tab:taxonomy} should not be read as mutually exclusive. A single plan may combine several of them. For example, exposing a service externally may require both exposure enlargement and persistent host modification; repairing a broken environment may combine unsafe dependency introduction with privilege expansion. What ties the taxonomy together is not the low-level API or syscall, but the type of boundary left under-specified and the corresponding completion pattern by which the agent turns that missing boundary into a host-side action sequence.

This distinction is important for two reasons. First, it clarifies why semantic-induced risks are systematic rather than accidental. The same classes of risky completion recur whenever the user states an end state without specifying the safe path. Second, it explains why simple action blocking is often insufficient. A policy that forbids one specific command may still leave the underlying completion pressure intact, causing the agent to find a different but equally risky path. In other words, the taxonomy is meant to classify how under-specified boundaries are operationalized, not just which final commands happen to be executed.

\section{Case Study and Qualitative Analysis}
We use OpenClaw as a representative case of an HAA that operates close to the host environment and therefore exposes the semantics-to-action gap in a particularly clear way. In the deployment model studied here, OpenClaw runs inside a Debian-based container while remaining coupled to persistent operator state through writable mounted directories. The point of this section is not to claim a newly discovered implementation bug in OpenClaw. Rather, it is to show that under-specified requests naturally admit multiple plans with materially different security costs even when the agent is containerized.

\subsection{Study Setup}
Our empirical evidence is trace-based and targeted. We use a live OpenClaw deployment in which the gateway and agent runtime execute inside a Debian/Bookworm container, while persistent configuration and workspace data remain writable through host-coupled mounts. We collect two families of traces. The first uses goal-only dry-run prompts against ordinary OpenClaw sessions to observe how the system completes generic requests such as environment setup, collaborator access, and deployment repair. The second uses a scoped project-local fixture inside the mounted workspace to test whether stronger locality cues push the model toward safer, more contained plans. The goal is not to estimate benchmark-scale success rates, but to characterize how semantic completion changes plan scope, persistence, and exposure.

We focus on traces for two reasons. First, traces are sufficient to reveal the system's preferred completion path under under-specified instructions. Second, they reduce the risk of accidental side effects while we are still characterizing the deployment boundary itself. In one fixture-based execution attempt, the runtime refused shell execution because the relevant exec hosts were unavailable in the current session. We treat that behavior as part of the observed deployment boundary rather than as a failure of the study design: it distinguishes what the runtime currently permits from what the model would otherwise choose to do.

Table~\ref{tab:cases} summarizes three representative scenarios.

\begin{table*}[t]
\centering
\caption{OpenClaw-centered scenarios illustrating security-divergent plans under the same goal.}
\label{tab:cases}
\footnotesize
\setlength{\tabcolsep}{5pt}
\begin{tabularx}{\textwidth}{L{2.1cm}L{2.8cm}YYL{2.5cm}}
\toprule
\textbf{Scenario} & \textbf{User goal} & \textbf{Relatively conservative plan} & \textbf{Riskier but still goal-valid plan} & \textbf{Primary risk classes} \\
\midrule
Environment setup & ``Get this project running on my machine.'' & Create a project-local virtual environment, inspect the dependency manifest, ask before system-wide changes, limit modifications to the workspace. & Install missing packages globally, modify shell startup files, change system interpreters, disable checks to bypass setup friction. & Privilege expansion, persistent modification, unsafe dependency introduction \\
\midrule
Service exposure & ``Make this app accessible to collaborators.'' & Use a scoped local reverse proxy or temporary authenticated tunnel, keep the service bound locally, surface the exposure decision explicitly. & Bind to \texttt{0.0.0.0}, open firewall ports broadly, disable authentication, expose a long-lived public endpoint by default. & Exposure enlargement, persistent modification \\
\midrule
Fault repair & ``Fix the deployment so it stops crashing.'' & Inspect logs, isolate the failing component, roll back only the implicated configuration, keep changes reversible. & Delete state aggressively, overwrite configs, relax permissions recursively, disable protective modules, restart multiple services forcefully. & Destructive repair, privilege expansion, sensitive-resource overreach. \\
\bottomrule
\end{tabularx}
\end{table*}

\begin{figure*}[t]
\centering
\includegraphics[width=0.8\textwidth]{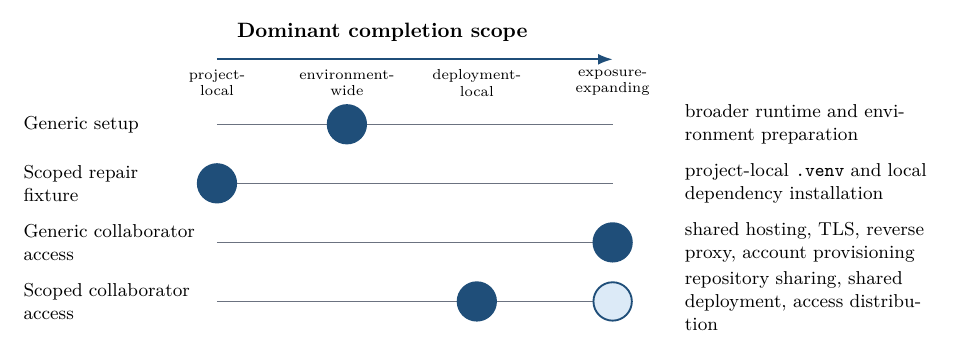}
\caption{Trace summary from the OpenClaw case study. More explicit locality cues narrow repair/setup plans, but exposure-oriented goals still push completion toward broader deployment and access distribution.}
\label{fig:prelim-matrix}
\end{figure*}

\subsection{Scenario A: Environment Setup}
The first scenario shows how a request such as ``get this project running on my machine'' can expand from workspace-local setup into broader environment reconfiguration. A comparatively conservative plan would inspect the project manifest, create an isolated environment, and treat system-wide or cross-workspace changes as exceptions that require justification or confirmation. A more aggressive plan could instead install packages globally inside the runtime, alter shell initialization, overwrite configuration, or write broadly into mounted workspace state. Both plans are goal-valid in the narrow sense that they aim to make the project run, but they differ sharply in persistence, scope, and side effects. The key issue is therefore not just whether the task succeeds, but whether the chosen completion path preserves the unstated boundary that the user likely intended: solve the project problem without broadly reconfiguring persistent state.

\subsection{Scenario B: Service Exposure}
The second scenario highlights the difference between accessibility and exposure. A request such as ``make this app accessible to my collaborators'' specifies a collaboration goal but leaves critical network-boundary semantics open. In practice, the agent may choose among several plausible paths: maintain local binding and create a temporary authenticated sharing path, expose a selected port to a restricted network segment, or make the service publicly reachable through a permissive binding or a third-party tunnel. In a containerized deployment, these choices remain security-relevant because the service boundary is bridged through published ports, host networking decisions, and persistent gateway state. A plan that quickly binds the service to a public interface, opens firewall rules, or weakens authentication may look efficient from a pure task-completion perspective, but it changes the system's exposure boundary in a way the user may never have intended to delegate.

\subsection{Scenario C: Fault Repair}
The third scenario illustrates why repair tasks are especially dangerous for HAAs. Requests such as ``fix the deployment so it stops crashing'' are semantically thin but operationally forceful: they emphasize recovery while leaving acceptable repair cost undefined. In an OpenClaw-like workflow, the agent may infer that deleting state, overwriting configuration files, widening permissions, disabling checks, or force-restarting services is justified if these actions quickly restore functionality. Such behavior is not irrational relative to the stated objective. It becomes problematic because the objective says nothing about reversibility, auditability, collateral effects, or the preservation of existing protections.

\subsection{Synthesis}
Taken together, the scenarios support the paper's central qualitative claim: the unsafe behavior is not best understood as an isolated bug or one-off misfire. It is a structural consequence of allowing a host-acting agent to optimize for task completion under incomplete safety semantics. The same pattern recurs across setup, exposure, and repair tasks because all three rely on goal-only instructions that leave the safe execution boundary implicit.

Our traces make this pattern concrete. For an under-specified request such as ``make this app accessible to my collaborators,'' repeated runs consistently completed the goal into a stronger exposure-oriented deployment path involving shared hosting, HTTPS, reverse proxying, and account provisioning. Those choices are semantically defensible, but they materially enlarge deployment scope and attack surface relative to the original request. By contrast, scoped fixture-based traces for a project-local repair target tended to prefer a project-local virtual environment and local dependency installation rather than broader environment-wide changes. However, when we kept the concrete project path but changed the goal back to collaborator access, the model again drifted toward repository publication, shared deployment, and access distribution. Figure~\ref{fig:prelim-matrix} summarizes these observations. These traces do not constitute a benchmark, but they support two linked claims: semantic completion can add security-significant structure that the user never explicitly delegated, and stronger locality cues narrow the risky plan space only when the goal itself does not continue to push toward wider deployment or exposure.

\section{Defense Design and Feasibility Discussion}
\subsection{Design Principles}
If the problem originates in semantic completion, then the defense should not begin only at the point of low-level action execution. The first design principle is therefore to \emph{separate goal specification from boundary specification}. Users should be able to express not only what outcome they want, but also whether persistence is allowed, whether internet exposure is acceptable, which resources are in scope, and when confirmation is required. Even coarse boundary fields are preferable to leaving all such constraints implicit. This principle directly targets the kinds of under-specified boundaries that recur across our taxonomy, especially persistence, exposure, scope, and recovery-cost ambiguity.

The second principle is \emph{risky-step elevation}. Plans often contain a small number of semantically pivotal actions such as using elevated privileges, exposing a service, introducing external dependencies, or making persistent changes. Those steps should be surfaced explicitly before execution, along with a brief rationale for why the agent believes they are needed. This converts hidden semantic completion into inspectable policy-relevant structure. In terms of the taxonomy, risky-step elevation is most naturally aimed at privilege expansion, exposure enlargement, unsafe dependency introduction, and destructive repair.

The third principle is \emph{plan transparency and auditability}. For HAAs, logging only the final commands is not enough. The system should retain the inferred goal decomposition, the candidate risky steps, and the reason a given path was chosen over alternatives. Such traces are useful not only for debugging and incident response, but also for improving future policy constraints and identifying recurrent under-specification patterns. This principle is especially important for cases such as fault repair and collaborator access, where the model can produce multiple semantically valid plans whose security costs differ sharply.

The fourth principle is \emph{execution-domain constraint}. Even with better semantic interfaces, some under-specification will remain unavoidable. Practical systems should therefore restrict the damage a poor completion can cause by running in constrained environments, limiting persistence by default, and elevating only narrowly scoped privileges when necessary. Anthropic's own computer-use guidance already recommends dedicated containers or virtual machines with minimal privileges for high-risk interaction settings \cite{ref_anthropic_computer_use}; the same logic applies to OpenClaw-like systems more broadly. Recent defense systems such as AgentSentinel, CSAgent, and AgentSys reinforce this direction through real-time auditing, context-aware access control, and explicit runtime state management for computer-use agents \cite{ref_agentsentinel,ref_csagent,ref_agentsys}. In our setting, this principle is what limits the operational impact of risky completion when semantic constraints are still incomplete.

Our traces also suggest that this principle matters semantically, not just operationally. When the task framing names a specific project path and keeps the objective local, the model is more likely to complete the request through project-local mechanisms such as a virtual environment instead of broader environment reconfiguration. This is a useful indication that better boundary specification can reduce risky completion pressure even before any low-level enforcement is applied.

\subsection{Conceptual Pipeline}
Taken together, these principles can be organized into a lightweight conceptual pipeline:
\begin{enumerate}
\item parse the user goal into an intended outcome and an explicit boundary profile;
\item annotate candidate plans for semantic risk classes such as exposure enlargement or destructive repair;
\item require elevation or confirmation for boundary-crossing steps;
\item execute within a constrained runtime; and
\item preserve an auditable trace of the goal-to-plan translation.
\end{enumerate}

This design remains intentionally lightweight. The point is not to propose a heavyweight prototype for this paper, but to show that the semantic framing yields actionable system implications. In practical terms, these mechanisms are compatible with existing agent loops, tool wrappers, and policy engines. The additional burden lies mainly in representing what is currently implicit: the boundaries that shape which goal-valid plans should be considered acceptable, and the small set of plan steps that most clearly signal completion pressure toward unsafe outcomes.

\subsection{Feasibility}
The feasibility claim in this paper is intentionally modest. The proposed direction does not require an agent to recover human intent perfectly or eliminate semantic ambiguity from natural-language interaction. Instead, it focuses on a smaller and more practical objective: making a limited set of high-impact boundary-crossing actions more visible, constrained, and reviewable before they silently become host-side effects.

This is plausible because many of the proposed mechanisms align with control points that already exist in OpenClaw-like systems. Goal-and-boundary separation can be introduced as lightweight structured request metadata. Risky-step elevation can be inserted between plan generation and tool execution. Plan transparency can be supported by extending existing traces or tool-call logs to retain boundary-relevant decisions rather than only final commands. Execution-domain constraint can build on mechanisms that many deployments already use, such as container confinement, limited mounts, scoped credentials, or stronger isolation for higher-risk tasks. In this sense, the proposal mainly requires making implicit safety assumptions explicit at existing decision points, rather than introducing a fundamentally new runtime architecture.

The design is also incrementally deployable. A system can begin by surfacing only a few high-impact actions, such as internet exposure, persistent host modification, or privilege elevation, and then gradually add coarse boundary profiles, lightweight audit traces, and stronger execution constraints. This staged approach is important for convenience-oriented agents: the goal is not to interrupt every ambiguous step, but to intervene selectively where semantic under-specification is most likely to produce disproportionate security cost.

For these reasons, the feasibility argument is not that semantic uncertainty can be solved completely, but that OpenClaw-like systems already provide enough architectural structure to support coarse boundary expression, selective risky-step elevation, lightweight auditing, and constrained execution in a practical and incremental manner.

\subsection{Limitations}
This study has several limitations. First, the empirical component is trace-based rather than benchmark-scale: our evidence is drawn from qualitative case analysis and execution traces rather than repeated evaluation across many models, tasks, and deployment settings. Second, the paper focuses on semantic under-specification as one important source of risk, not the only one. Real deployments are also affected by prompt injection, over-privileged tools, implementation flaws, operator misconfiguration, and model-specific weaknesses. Third, our OpenClaw experiments are conducted in a Debian/Bookworm containerized deployment with writable host-coupled state rather than a fully host-native bare-metal setup. We regard this as a meaningful and security-relevant deployment model, but the exact operational consequences can differ across deployment architectures. Finally, the proposed taxonomy is an organizing framework rather than a complete ontology of all possible agent-induced risks.

\subsection{Ethical Considerations and Disclosure}
The purpose of this paper is defensive: to clarify how convenience-oriented agent interaction can lead to risky host-side plan completion and to motivate safer system design. We therefore avoid presenting exploit chains, persistence tricks, or step-by-step abuse guidance aimed at compromising third-party systems. The OpenClaw-centered analysis in this paper is used as a representative case for understanding semantic-induced risk, not as a claim of a newly discovered implementation vulnerability in the project. When experiments involve live agent runtimes, they should be conducted only in controlled environments owned or authorized by the researcher, with clear attention to mounted state, exposed services, and accidental persistence. This is especially important for agent systems whose memory and stored context may themselves become security-relevant control surfaces \cite{ref_amemguard,ref_zombie_agents,ref_storage_to_steering}. If future work uncovers concrete implementation bugs beyond the semantic issues analyzed here, such findings should be handled through responsible disclosure to the relevant maintainers.

\section{Conclusion}
This paper argued that security risks in host-acting agents cannot be understood solely as problems of excessive permissions, implementation flaws, or adversarial prompt manipulation. A more fundamental issue lies in semantic under-specification: users specify desired outcomes while leaving execution boundaries, persistence, and exposure expectations implicit, and the resulting completion process gives agents room to select security-divergent plans before execution begins.

Taken together, the semantic threat model, taxonomy, and OpenClaw-centered study in this paper support one central claim: agent security must govern not only which actions are permitted, but also how convenience-oriented systems translate goal-only instructions into executable plans. In this sense, the relevant security boundary does not begin at execution alone. It begins earlier, at the stage where missing task semantics are completed into concrete operational choices.

This paper adopts a bounded scope. It does not claim that semantic under-specification is the only cause of unsafe agent behavior, nor that the taxonomy is complete, nor that we have already provided a full production defense. Instead, the contribution is to isolate an under-theorized source of risk, show why it is structurally different from adversarial hijacking alone, and provide a framework that can guide stronger measurement and mitigation in future work.

\end{document}